\documentclass[prl,twocolumn,superscriptaddress]{revtex4-2} 
\usepackage{mathrsfs}
\usepackage{amsfonts}
\usepackage{amsmath,amssymb}
\usepackage{amsthm}
\usepackage{txfonts}
\usepackage{amssymb}
\usepackage{graphicx,subfigure}
\usepackage{bm}
\usepackage{color}
\usepackage[normalem]{ulem}
\usepackage{dcolumn} 
\usepackage{bbold}
\usepackage[table]{xcolor}
\usepackage{float}
\usepackage{tabularx}
\usepackage{hyperref}
\newcommand{\ket}[1]{|#1\rangle}
\newcommand{\bra}[1]{\langle #1|}
\newcommand{\Tr}{\text{Tr}}

\begin{document}

\title{Quantum analog of Landau-Lifshitz-Gilbert dynamics} 

\author{Yuefei Liu}
\thanks{These two authors contributed equally.}
\affiliation{Department of Applied Physics, School of Engineering Sciences, 
KTH Royal Institute of Technology, AlbaNova University Center, 
SE-10691 Stockholm, Sweden}
\email{yuefei@kth.se}

\author{Ivan P. Miranda}
\thanks{These two authors contributed equally.}
\affiliation{Department of Physics and Astronomy, Uppsala University, 
Box 516, SE-751 20 Uppsala, Sweden}
\email{ivan.miranda@alumni.usp.br}
\affiliation{Department of Physics and Electrical Engineering, Linnaeus University, 
SE-39231 Kalmar, Sweden}

\author{Lee Johnson}
\affiliation{Department of Physics and Astronomy, Uppsala University, Box 516, 
SE-751 20 Uppsala, Sweden}

\author{Anders Bergman}
\affiliation{Department of Physics and Astronomy, Uppsala University, Box 516, 
SE-751 20 Uppsala, Sweden}

\author{Anna Delin}
\affiliation{Department of Applied Physics, School of Engineering Sciences, 
KTH Royal Institute of Technology, AlbaNova University Center, 
SE-10691 Stockholm, Sweden}
\affiliation{Swedish e-Science Research Center (SeRC), 
KTH Royal Institute of Technology, SE-10044 Stockholm, Sweden}
\affiliation{Wallenberg Initiative Materials Science for Sustainability (WISE), 
KTH Royal Institute of Technology, SE-10044 Stockholm, Sweden}

\author{Danny Thonig}
\affiliation{School of Science and Technology, \"Orebro University, 
SE-701 82, \"Orebro, Sweden}
\affiliation{Department of Physics and Astronomy, Uppsala University, 
Box 516, SE-751 20 Uppsala, Sweden}

\author{Manuel Pereiro}
\affiliation{Department of Physics and Astronomy, Uppsala University, 
Box 516, SE-751 20 Uppsala, Sweden}

\author{Olle Eriksson}
\affiliation{Department of Physics and Astronomy, Uppsala University, 
Box 516, SE-751 20 Uppsala, Sweden}
\affiliation{WISE-Wallenberg Initiative Materials Science, Uppsala University, 
Box 516, SE-751 20 Uppsala, Sweden}

\author{Vahid Azimi-Mousolou}
\affiliation{Department of Applied Mathematics and Computer Science, 
Faculty of Mathematics and Statistics, University of Isfahan, 
Isfahan 81746-73441, Iran}
\affiliation{Department of Physics and Astronomy, Uppsala University, 
Box 516, SE-751 20 Uppsala, Sweden}

\author{Erik Sj\"oqvist}
\affiliation{Department of Physics and Astronomy, Uppsala University, 
Box 516, SE-751 20 Uppsala, Sweden}
\email{erik.sjoqvist@physics.uu.se}

\date{\today}

\begin{abstract}
The Landau-Lifshitz-Gilbert (LLG) and Landau-Lifshitz (LL) equations play an essential 
role for describing the dynamics of magnetization in solids. While a quantum analog of 
the LL dynamics has been proposed in [Phys.~Rev.~Lett.~{\bf 110}, 147201 (2013)], the 
corresponding quantum version of LLG remains unknown. Here, we propose such a 
quantum LLG equation that inherently conserves purity of the quantum state. We 
examine the quantum LLG dynamics of a dimer consisting of two interacting 
spin-$\frac{1}{2}$ particles. Our analysis reveals that, in the case of ferromagnetic 
coupling, the evolution of initially uncorrelated spins mirrors the classical LLG dynamics. 
However, in the antiferromagnetic scenario, we observe pronounced deviations from 
classical behavior, underscoring the unique dynamics of becoming a spinless state,  
which is non-locally correlated. Moreover, when considering spins that are initially 
entangled, our study uncovers an unusual form of revival-type quantum correlation 
dynamics, which differs significantly from what is typically seen in open quantum 
systems.
\end{abstract}

\maketitle

\textit{Introduction.---} The Landau-Lifshitz-Gilbert (LLG) \cite{gilbert04} and 
Landau-Lifshitz (LL) \cite{landau35} equations describe the dynamics of magnetization 
in solids on an atomistic, classical level \cite{eriksson17}. Several publications 
have used this approach to describe magnetization dynamics of 
topological objects \cite{pereiro14}, the demagnetization of fcc Ni in pump probe 
experiments \cite{evans15,pankratova22}, and the magnetization reversal of ferrimagnetic 
FeGd alloys \cite{chimata15}. While the LL and LLG equations treat the magnetization 
as a classical vector, the underlying degrees of freedom are quantum spins. This raises 
the question whether quantum versions of these equations exist. In the LL case, this 
problem has been addressed by Wieser \cite{wieser13,wieser15,wieser16}, while the 
quantum analog of LLG remains unknown. 

Here, we extend Wieser's work and propose a quantum analog of the LLG equation. 
This quantum LLG (q-LLG) equation describes the dynamics of the density operator 
of quantum spins. We examine similarities and differences between the resulting 
quantum and classical dynamics for a dimer consisting of two spin-$\frac{1}{2}$ particles. 
We show that the proposed q-LLG generally differs from Wieser's quantum LL (q-LL) equation 
for multi-spin systems in non-pure states, while the two equations may be equivalent up 
to a rescaling of time in the single spin case, similar to the relation between their classical 
counterparts. 

We first describe the classical LLG and LL dynamics. Consider a 
system of interacting magnetic moments ${\bf m}_k$, being exposed to an external 
magnetic field ${\bf B}$. The LLG and LL equations both  describe damped precession 
of the ${\bf m}_k$:s around their local effective magnetic field 
${\bf B}_k = - \frac{\partial H}{\partial {\bf m}_k}$. These effective fields contain 
contributions from ${\bf B}$ as well as from all ${\bf m}_{l\neq k}$ via the magnetic 
many-body Hamiltonian $H$, which 
typically include Heisenberg and Dzyaloshinskii-Moriya terms, i.e., 
$H=-\sum_k {\bf B} \cdot {\bf m}_k + \mu_B^{-2} \sum_{l < k} \left[ 
J_{lk}{\bf m}_{l} \cdot {\bf m}_{k} + {\bf D}_{kl} \cdot \left({\bf m}_k \times {\bf m}_l \right) \right]$ 
with $\mu_B$ the Bohr magneton.

The LLG equations read 
\begin{eqnarray}
\dot{\bf m}_k = \gamma_g {\bf m}_k \times {\bf B}_k - \frac{\alpha}{\left| {\bf m}_k \right|} 
{\bf m}_k \times \dot{{\bf m}}_k ,
\label{eq:llg}
\end{eqnarray}  
 with $\gamma_g$ the gyromagnetic ratio and $\alpha$ the dimensionless Gilbert damping. 
 Note that most derivations of Eq.~\eqref{eq:llg} assume that $\left| {\bf m}_k \right|$ is a 
 constant of the motion. This latter fact can be used in Eq.~\eqref{eq:llg} to find the LL 
 equations (see, e.g., Ref.~\cite{eriksson17})
\begin{eqnarray}
\dot{\bf m}_k = \tilde{\gamma}_g {\bf m}_k \times {\bf B}_k - 
\frac{\lambda}{\left| {\bf m}_k  \right|} {\bf m}_k\times \left( {\bf m}_k \times {\bf B}_k \right) 
\label{eq:ll}
\end{eqnarray}
with the rescaled gyromagnetic ration $\tilde{\gamma}_g = \gamma_g/(1+\alpha^2)$ 
and LL damping rate $\lambda = \alpha \gamma_g/(1+\alpha^2)$. The way the 
LL equation can be derived from the LLG equation demonstrates that they are 
equivalent up to a rescaling of time $t \mapsto (1+\alpha^2)t$ \cite{lakshmanan84}. 

\textit{Master equation and properties.---} Inspired by the above classical formulation, 
we propose the trace preserving q-LLG analog as 
\begin{eqnarray}
\dot{\varrho} = \frac{i}{\hbar} [\varrho,H] + i \kappa [\varrho,\dot{\varrho}] 
\label{eq:qllg}
\end{eqnarray}
with $\varrho$ the density operator. 
The term $i \kappa [\varrho,\dot{\varrho}]$ that modifies the Liouville-von 
Neumann equation has a \textit{dissipative} (damping-like) character with $\kappa$ 
the  dimensionless damping rate. In addition to the trace, Eq.~\eqref{eq:qllg} 
can be shown to preserve Hermiticity and non-negativity of $\varrho (t)$ \cite{sm}.

The first property that can be immediately verified from Eq.~\eqref{eq:qllg} is the purity 
conservation, $\frac{d}{dt} \Tr{\varrho^2} = 0$. This implies a fundamental difference 
from the well-known Lindbladian superoperator \cite{Breuer2007}, which also acts as 
a dissipator, but imposes to the system a much less strict condition, for which the purity 
is not generally conserved \cite{Manzano2020}. Thus, in the case of q-LLG, the purity 
of the density operator becomes the quantum analog of the classical magnetization. 

The conservation of purity seems to suggest similar results for different entropy 
measures. Indeed, one can observe that the Rényi entropy $S_{\delta} (\varrho) = 
(1-\delta)^{-1} \ln \Tr \varrho^{\delta}$, $\delta \neq 1$, which tends asymptotically 
to the von Neumann entropy $S=-\Tr(\varrho\ln\varrho)$ when $\delta \rightarrow 1$, 
is conserved. Thus, although Eq.\,(\ref{eq:qllg}) includes a 
damping-like term, the conserved purity ensures that there will be no intrinsic loss of 
information during the damped dynamics.   

\textit{Connection to classical LLG and q-LL.---}
To illustrate the connection of q-LLG with the classical LLG, let us consider 
the simplest case of a single spin-$s$ particle. Suppose the evolving state of the 
particle takes the form 
\begin{eqnarray}
\varrho = \frac{1}{2s+1} \left( \mathbb{1} + \boldsymbol{\eta} \cdot {\bf S} \right)     
\label{eq:spinstate}
\end{eqnarray}
with the spin operator ${\bf S}$ and identity $\mathbb{1}$ acting on the $2s+1$ 
dimensional Hilbert space $\mathcal{H}$ associated with the spin state. We 
wish to interpret $\boldsymbol{\eta}$ as the quantum analog of the classical 
magnetization discussed above. This implies that $|\boldsymbol{\eta}|$ should 
be a constant of the motion and $\boldsymbol{\eta}$ should satisfy a classical 
LLG-type equation \cite{remark1}. 

To see that $|\boldsymbol{\eta}|$ is constant in time, we note that 
\begin{eqnarray}
\hbar^2 |\boldsymbol{\eta}|^2 & = & \frac{3(2s+1)}{s(s+1)} \left( \Tr \varrho^2 
- \frac{1}{2s+1} \right) ,   
\label{eq:qm}
\end{eqnarray}
from which immediately follows $\frac{d}{dt} |\boldsymbol{\eta}|^2 \propto \frac{d}{dt} 
\Tr \varrho^2 = 0$.  To demonstrate that $\boldsymbol{\eta}$ satisfies a classical 
LLG-type equation, we consider the case where the spin is exposed to a magnetic 
field ${\bf B}$ as described by a Zeeman Hamiltonian $H = -\gamma_g {\bf B} \cdot {\bf S}$. 
By inserting Eq.~\eqref{eq:spinstate} and $H$ into Eq.~\eqref{eq:qllg}, one finds 
\begin{eqnarray}
\dot{\boldsymbol{\eta}} = \gamma_g \boldsymbol{\eta} 
\times {\bf B} - \frac{\kappa \hbar}{2s+1} \boldsymbol{\eta} 
\times \dot{\boldsymbol{\eta}} , 
\label{eq:qllg2}
\end{eqnarray}
where the spin commutation relation $[{\bf a} \cdot {\bf S},{\bf b} \cdot {\bf S}] = 
i\hbar( {\bf a} \times {\bf b}) \cdot {\bf S}$ has been used. Clearly, Eq.~\eqref{eq:qllg2} 
is identical to Eq.~\eqref{eq:llg} and in this single spin case we find that 
$\kappa \hbar |\boldsymbol{\eta}| /(2s+1)$ is the dimensionless quantum analog 
of the Gilbert damping. 

Next, we  argue that the proposed q-LLG equation is generally different from 
the q-LL form $\dot{\varrho} = \frac{i}{\hbar}[\varrho,H] - \frac{\kappa}{\hbar} 
[\varrho,[\varrho,H]]$ found by Wieser \cite{wieser13}. To see this, we rewrite 
Eq.~\eqref{eq:qllg} as 
$\dot{\varrho} + \kappa^2 [\varrho,[\varrho,\dot{\varrho}]] = \frac{i}{\hbar}[\varrho,H] - 
\frac{\kappa}{\hbar} [\varrho,[\varrho,H]]$, 
which is the desired q-LL form if and only if the left-hand side is proportional to 
$\dot{\varrho}$. This is the case for general pure states ($\varrho^2 = \varrho$) 
and for a single spin in a non-pure state of the form Eq.~\eqref{eq:spinstate}. 
Indeed, one finds for these two cases 
$\dot{\varrho} + \kappa^2 [\varrho,[\varrho,\dot{\varrho}]] = (1+\kappa^2) \dot{\varrho}$
and $\dot{\varrho} + \kappa^2 [\varrho,[\varrho,\dot{\varrho}]] = 
\left( 1 + \kappa^2  \hbar^2 |\boldsymbol{\eta}|^2/(2s+1)^2 \right) \dot{\varrho}$, respectively.   
On the other hand, in other cases, no such simplification is generally possible, and 
the q-LLG and q-LL dynamics are therefore generally different  \cite{remark2}.  

While one can convince oneself that the q-LLG proposed in Eq.~\eqref{eq:qllg} 
is the only possible form of master equation that exactly reproduces the classical 
description for uncoupled magnetic systems, this equivalence is no longer the 
case when coupling is included \cite{remark3}. To see this, it is sufficient to compare the number 
of dynamical variables for $N$ particles in LLG and q-LLG. For spin-$\frac{1}{2}$, 
this number is $4^N-1$ \cite{remark4}, while it is $3N$ in the classical treatment. 
Thus, the classical-like variables associated with the reduced density operators 
for the individual quantum spins is only a small portion of the total number of 
quantum mechanical degrees of freedom. 
To illustrate this extra richness, we shall in the following examine the q-LLG 
dynamics of a dimer consisting of two spin-$\frac{1}{2}$ particles.

\textit{q-LLG dynamics of a spin dimer.---} The spin dimer is described by the 
tensor state space $\mathcal{H}^{\otimes 2}$ and locally by the dimensionless 
Bloch vectors ${\bf r}_k = \frac{\hbar}{2} \boldsymbol{\eta}_k$ associated with 
the single spin reduced density operators with $k=1,2$ labelling the two spins. 
A general state of the dimer can be written as 
\begin{eqnarray}
\varrho = \frac{1}{4}  T_{\alpha \beta} \sigma_{\alpha} \otimes \sigma_{\beta}, 
\label{eq:rho_general}
\end{eqnarray}
where $\sigma_{\alpha} \otimes \sigma_{\beta}$ is a member of the Pauli 
group $\mathscr{P}_2 = \left\{\sigma_0 \equiv 
\mathbb{1},\sigma_x,\sigma_y,\sigma_z \right\}^{\otimes 2} \times \{\pm 1,\pm i\}$ and 
the Einstein summation convention is used. We use Greek and Latin indices to run 
over $\{ 0, x, y, z \}$ and $\{ x, y, z \}$, respectively.  $T_{\alpha \beta}$ are elements 
of a $4 \times 4$ matrix $T$, for which $T_{00} = 1$, $T_{k0} =  r_{1;k}$, and $T_{0l} = r_{2;l}$,
where the first constraint ensures normalization and the latter two expressions give the 
components of the Bloch vectors ${\bf r}_1$ and ${\bf r}_2$ of the two spins. The remaining 
$T_{kl}$ form the $3\times 3$ correlation matrix.

The violation of Bell inequalities for certain quantum states gives a precise notion of 
non-classical correlations, as such $\varrho$ does not admit a local classical description 
\cite{werner89}. To emphasize the non-classicality of the dynamics in the simulations 
below, we therefore use Bell non-locality $B(\varrho)$ \cite{horst13} as our preferred 
correlation measure. $B(\varrho)$ is defined in terms of the measurement setting that 
maximizes the violation of the CHSH inequality \cite{clauser69} for a given $\varrho$. 
As shown in Ref.~\cite{horodecki95}, this maximum is given by the two largest 
singular values $u_1$ and $u_2$ of $T_{kl}$ in the sense that there exists a 
measurement setting that violates CHSH if and only if $1<u_1^2+u_2^2 \leq 2$, 
where the upper equality corresponds to the Cirel'son bound of maximal violation 
\cite{cirelson80}. This suggests the measure \cite{horst13} 
$B(\varrho) = \sqrt{\max \left\{u_1^2+u_2^2-1,0 \right\}}$ 
of Bell non-locality. A given state $\varrho$ contains non-classical correlations 
if and only if $B(\varrho) \neq 0$.   

\begin{figure}[b]
\centering
\includegraphics[width=240pt]{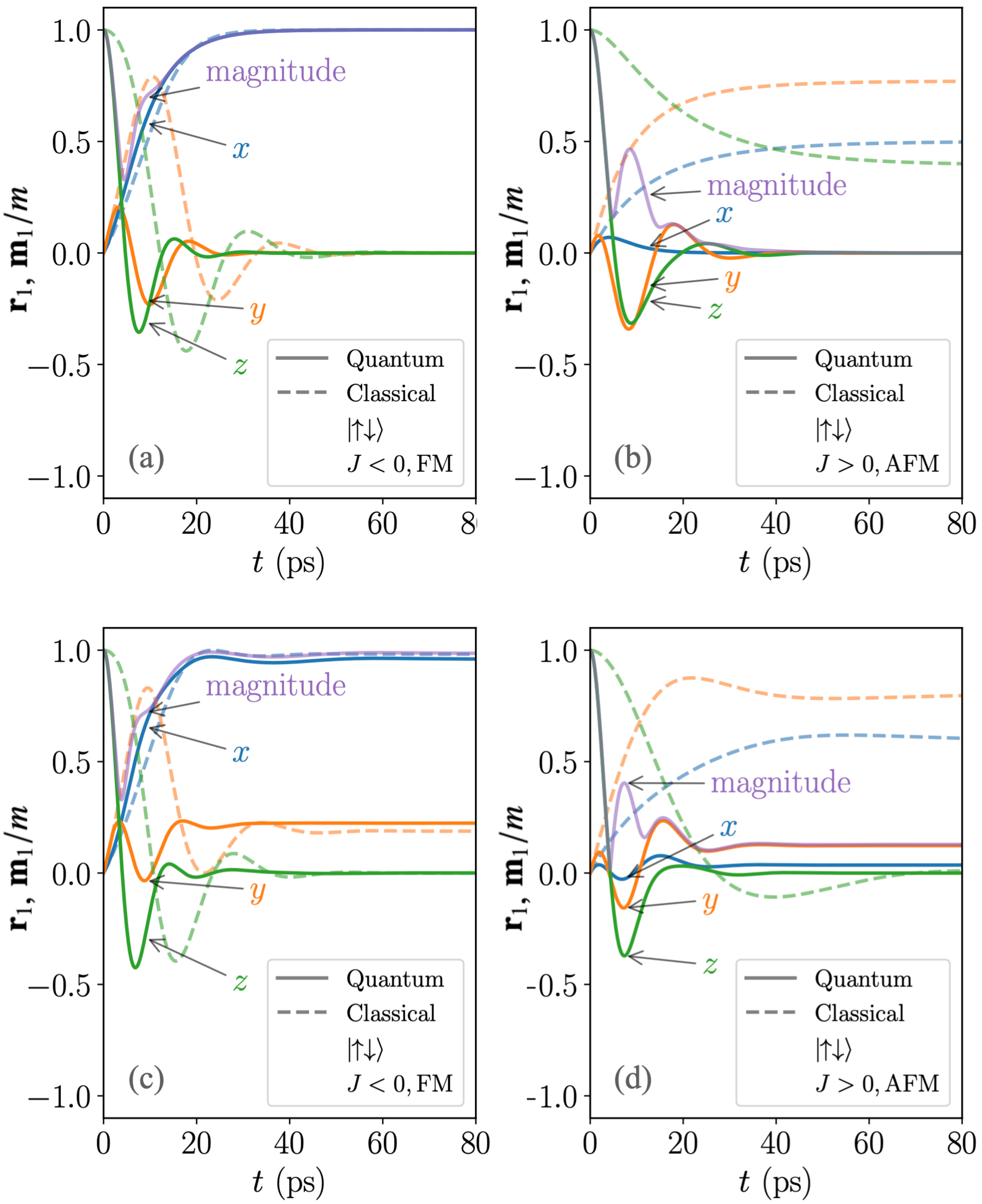}
\vspace{-4mm}
\caption{(Color online) q-LLG and LLG dynamics of a dimer initially prepared in the 
AFM-type product state $\ket{\!\!\uparrow \downarrow}$. The quantum dynamics is 
illustrated by the Bloch vector (solid lines) and the classical counterpart by the 
magnetization (dashed lines). (a) and (b) show the dynamics for FM ($J<0$) and 
AFM ($J>0$) exchange coupling, respectively, with in absence of DMI ($D=0$); 
(c) and (d) are the corresponding plots for $D/|J| = 0.6$. Due to symmetry, only 
${\bf r}_1$ (see text for definition) and ${\bf m}_1/m$ are shown. Physical 
parameters $B_0 = 1.00 \ {\rm T}$, $|J|/B_0=1\mu_B$, and $\kappa = \alpha=0.5$ 
are used.}
\vspace{-2mm}
\label{fig:llg_qllg}
\end{figure}

We now consider the dimer Hamiltonian
\begin{eqnarray}
H  =  -\gamma_g {\bf B} \cdot \left( {\bf S}_{1} + {\bf S}_{2} \right) + 
\frac{4J}{\hbar^2} {\bf S}_1 \cdot {\bf S}_2 + 
\frac{4{\bf D}}{\hbar^2} \cdot \left( {\bf S}_1 \times {\bf S}_2 \right),
\label{eq:H1}
\end{eqnarray}
with ${\bf B}$ an external magnetic field, $J$ the Heisenberg exchange coupling 
strength, and the Dzyaloshinskii–Moriya interaction (DMI) term with 
${\bf D} = D(0,0,1)$ ($D$ is the DMI strength). We take $\kappa = \alpha = 0.5$ 
to magnify the effects of the non-linear terms in the q-LLG and LLG equations; 
this choice is also physically consistent with the fact that Gilbert damping increases 
in low dimensions \cite{Steiauf2005}. Furthermore, we assume spin-$\frac{1}{2}$ 
and use the standard basis $\sigma_z \ket{\!\!\uparrow} = +\ket{\!\!\uparrow}$ 
and $\sigma_z \ket{\!\!\downarrow} = -\ket{\!\!\downarrow}$. 

First, we compare the q-LLG and LLG dynamics by examining the Bloch vectors 
${\bf r}_1$ and ${\bf r}_2$ of the reduced states $\varrho_1 = {\rm Tr}_2 \varrho$ 
and $\varrho_2 = {\rm Tr}_1 \varrho$, respectively, and the classical magnetizations 
${\bf m}_1$ and ${\bf m}_2$. We consider an antiferromagnetic (AFM) initial state 
of q-LLG dynamics defined as $\ket{\!\!\uparrow\downarrow}$ and corresponding 
initial magnetization ${\bf m}_1 = -{\bf m}_2 = m(0, 0, 1)$. We examine the resulting 
dynamics for ferromagnetic (FM, $J<0$) and AFM ($J>0$) exchange coupling. We 
use ${\bf B} = B_0 (1,0,0)$, i.e., an external field in the $x$ direction, and assume 
that the splitting due to the Zeeman and Heisenberg exchange is of the same order 
of magnitude by taking $m = J/B_0 = 1\mu_B = 6.58 \times 10^{-2} \ {\rm meV/T}$ 
and $B_0 = 1.00 \ {\rm T}$, which corresponds to the time scale 
$(\gamma_g B_0)^{-1} \sim 10 \ {\rm ps}$. Spins in this weak Heisenberg coupling 
regime can be realized in, e.g., quantum dots (see Ref.~\cite{trauzettel07}).

Figure \ref{fig:llg_qllg} displays the q-LLG and LLG dynamics of the dimer. We show 
only the evolution of one of the subsystems, since the other behaves in a similar 
fashion. A striking general feature of the q-LLG dynamics can be noticed: while 
the purity is conserved, the length of the Bloch vector is not generally preserved, 
even with this simple input state. This can be seen by comparing Figs.~\ref{fig:llg_qllg} 
(a) and (b). For an FM exchange, the quantum description is very similar to that 
of the classical simulation, and both approaches a fully saturated moment pointing 
along the $x$ direction. However, as shown in Fig.~\ref{fig:llg_qllg} (b), an AFM-type 
exchange interaction can lead to that the length of the Bloch vector is completely 
quenched by the quantum dynamics, effectively becoming a \textit{spinless} state; 
a behavior that is impossible in the classical description. One may further notice 
smaller deviations of the two approaches, e.g., that in Fig.~\ref{fig:llg_qllg} (a) 
the q-LLG is qualitatively similar to LLG results, with the discernible difference 
that the dynamics of the q-LLG equation is faster than that of the LLG equation. 
The effect of a non-zero DMI is shown in Figs.~\ref{fig:llg_qllg} (c) and (d). While 
the qualitative similarity between q-LLG and LLG remains for FM exchange 
coupling, the asymptotic quantum state is no longer fully quenched, i.e., the Bloch 
vector has a small but finite magnitude in the AFM case. The latter can be seen 
as an effect of the singlet-triplet coupling induced by the DMI. 

\begin{figure}[t]
\centering
\includegraphics[width=220pt]{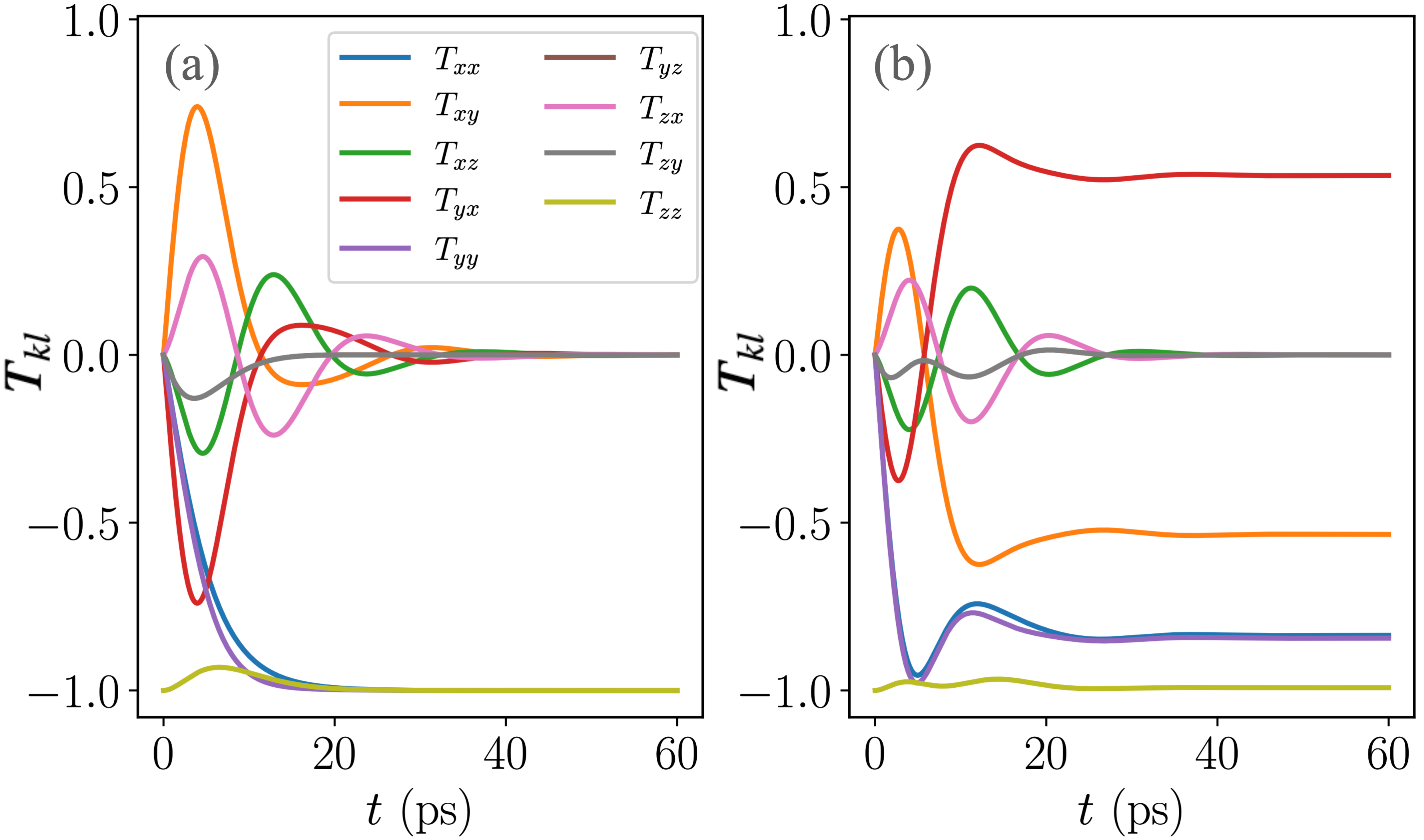}    
\vspace{-4mm}
\caption{(Color online) The correlation matrix elements $T_{kl}$ with (a) $D/|J|=0$ 
and (b) $D/|J|=0.6$, driven by q-LLG dynamics for an AFM ($J=1$) coupled dimer 
with initial product AFM-type state $\ket{\!\!\uparrow \downarrow}$. Physical 
parameters $B_0 = 1.00 \ {\rm T}$, $|J|/B_0=1\mu_B$, and $\kappa=0.5$ are used.}
\label{fig:qllg_T_d.pdf}
\end{figure}

To better understand the behavior depicted in Fig.~\ref{fig:llg_qllg}, we next turn to 
the correlation matrix $T_{kl}$. We again choose an initial AFM-type product state 
$\ket{\!\!\uparrow\downarrow}$ and focus on AFM exchange coupling $J>0$. Figure 
\ref{fig:qllg_T_d.pdf} shows the $T_{kl}$:s as a function of time. In 
Fig.~\ref{fig:qllg_T_d.pdf} (a), we see that $T_{kl}$ tends asymptotically to 
${\rm diag} \{ -1,-1,-1 \}$, which corresponds to the singlet Bell state 
$\ket{\Psi_-} = \frac{1}{\sqrt{2}} ( \ket{\!\! \uparrow \downarrow} - \ket{\!\! \downarrow \uparrow})$. 
This explains the steady state seen in Fig.~\ref{fig:llg_qllg} (b), as the reduced 
states of a Bell state have vanishing Bloch vector. In other words, while the Bloch 
vectors become completely quenched, information has been transferred to the 
non-local correlation of the spins in the steady state limit, so as to conserve the 
entropy during the process. Furthermore, the numerical results shown in 
Fig.~\ref{fig:qllg_T_d.pdf} (b) demonstrate the effect of a non-zero DMI on 
the q-LLG dynamics of the correlation matrix elements $T_{kl}$. The effect 
shows up as a symmetric splitting of the off-diagonal pair $T_{xy}$ and 
$T_{yx}$. This splitting effect is the explanation why ${\bf r}_1$ shown in 
Fig.~\ref{fig:llg_qllg} (d) no longer tends to a state with vanishing Bloch vector. 

\begin{figure}[h]
\centering
\includegraphics[width=200pt]{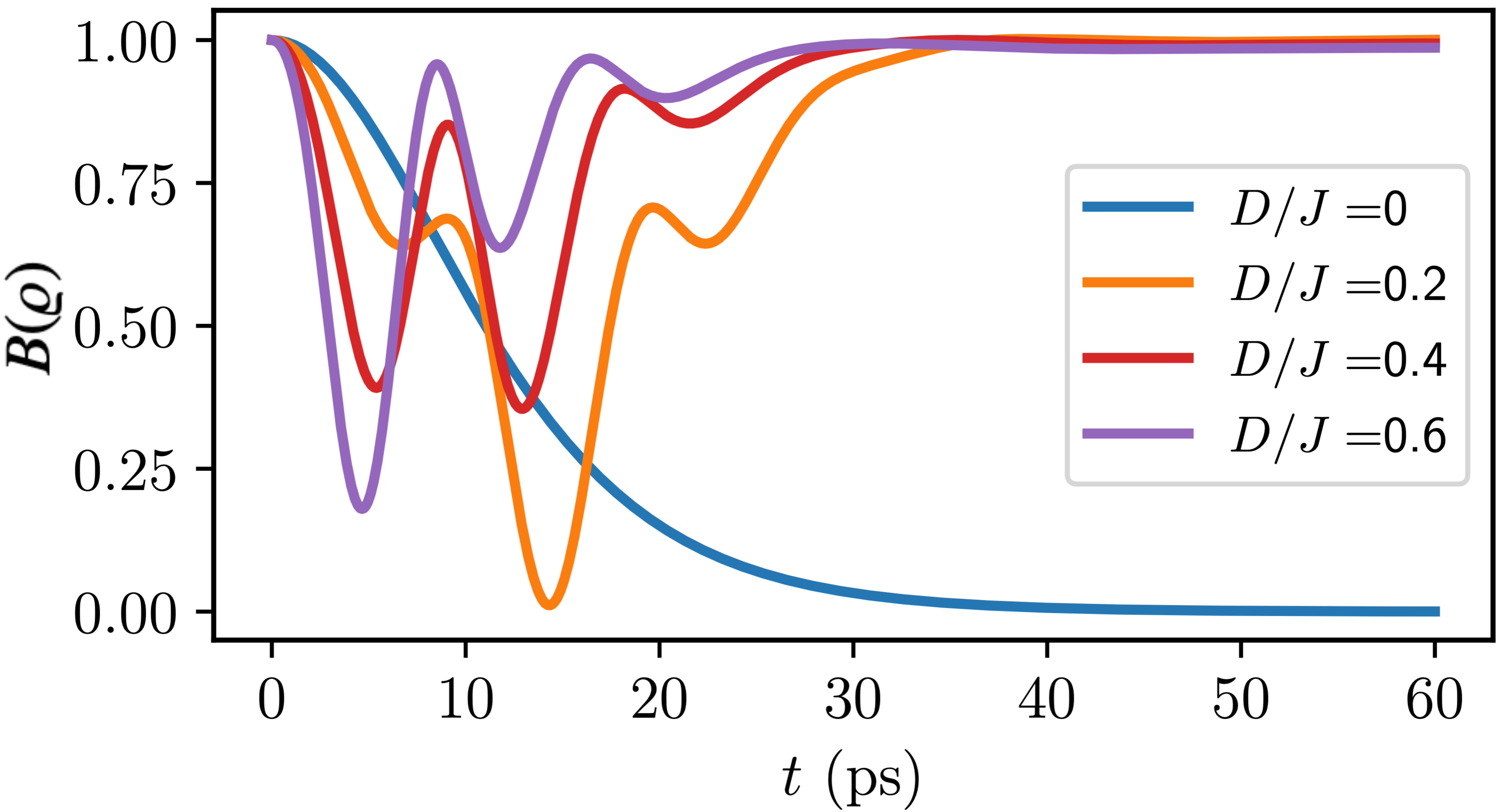} 
\vspace{-4mm}
\caption{(Color online) Quantum non-locality driven by the q-LLG dynamics 
with zero and non-zero DMI strength, in an AFM ($J>0$) exchange coupling 
dimer. We use ${\bf D} \perp {\bf B}$ and the initial state taken to be the Bell 
state $\ket{\Psi_+} = \frac{1}{\sqrt{2}}(\ket{\!\!\uparrow \downarrow} + 
\ket{\!\!\downarrow \uparrow})$. Physical parameters $B_0 = 1.00 \ {\rm T}$, 
$|J|/B_0=1\mu_B$, and $\kappa = 0.5$ are used.}
\label{fig:QLLG_Ds}
\end{figure}

We next consider the effect of the DMI term for initially correlated quantum spins. 
An AFM coupled dimer with an initial Bell state $\ket{\Psi_+}= \frac{1}{\sqrt{2}} 
(\ket{\!\!\uparrow \downarrow} + \ket{\!\!\downarrow \uparrow})$ is considered. 
In Fig.~\ref{fig:QLLG_Ds}, non-locality is shown for different DMI strength $D$. 
After an intermediate oscillatory phase, the system evolves asymptotically to 
a maximally non-local steady state for $D\neq 0$, while it decays monotonically 
to zero when $D=0$. These results can be understood as follows. In absence 
of DMI, the singlet Bell state $\ket{\Psi_-}$ is decoupled from the triplet states, 
which means that the system, initially in the triplet Bell state $\ket{\Psi_+}$, will 
remain in the triplet subspace. The dissipative dynamics thereby forces the spins 
to approach the FM-type product state $\frac{1}{2} \left( \ket{\!\!\uparrow} + 
\ket{\!\!\downarrow} \right)^{\otimes 2}$, which is the lowest energy eigenstate 
in the triplet subspace. In contrast, a non-zero $D$ couples the singlet and 
triplet states, thereby opening up a route towards the energetically favorable 
$\ket{\Psi_-}$, which is maximally non-local. 

\begin{figure}[h]
\centering
\includegraphics[width=230pt]{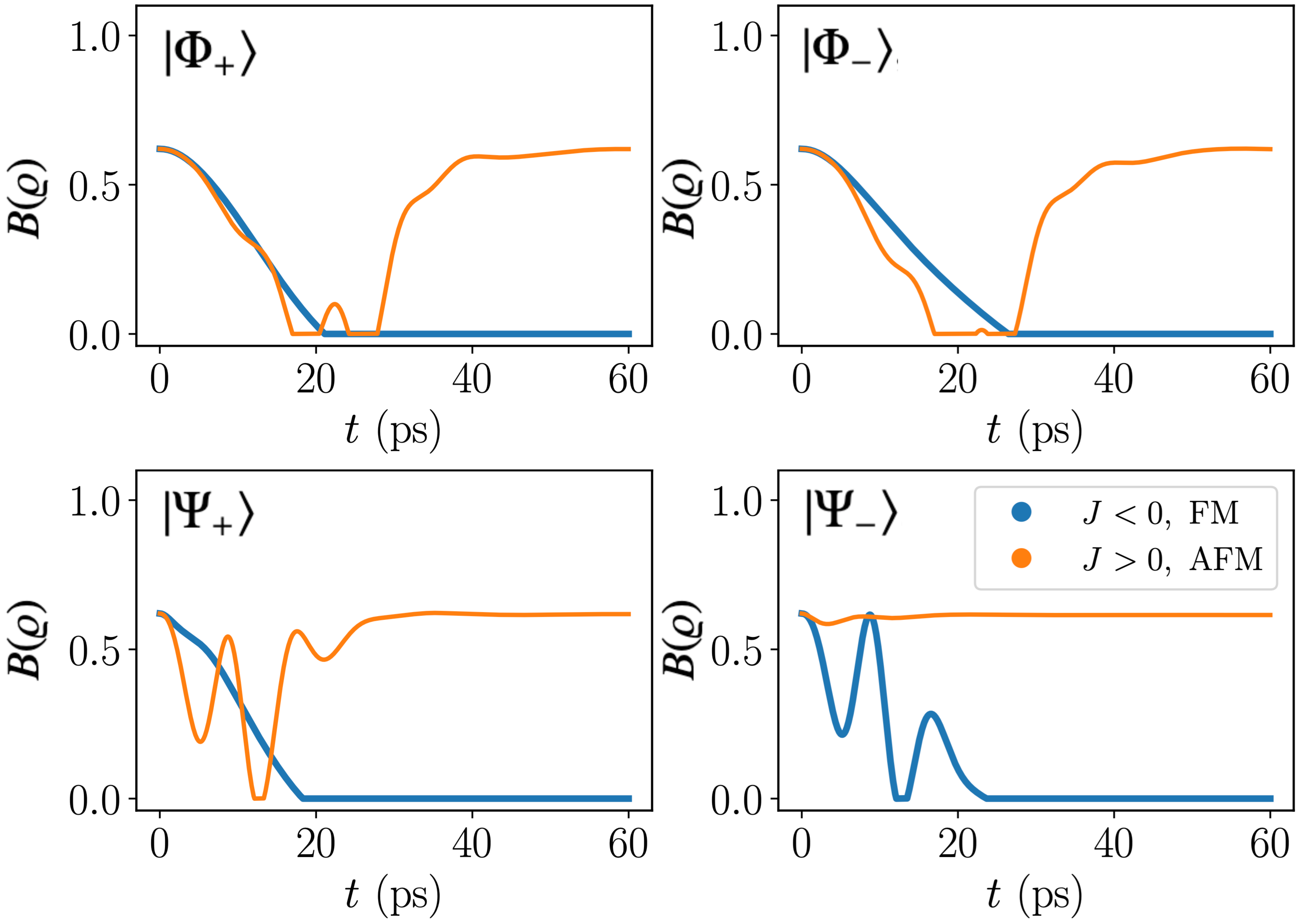}
\vspace{-4mm}
\caption{(Color online) q-LLG dynamics of non-locality for initial Werner states 
$\varrho_W = \frac{1-p}{4} \mathbb{1} + p\ket{\psi}\bra{\psi}$ with $\ket{\psi}$ 
being the Bell states $\ket{\Phi_{\pm}} = \frac{1}{\sqrt{2}} \left( \ket{\!\!\uparrow\uparrow} 
\pm \ket{\!\!\downarrow\downarrow} \right)$ and $\ket{\Psi_{\pm}} = 
\frac{1}{\sqrt{2}} \left( \ket{\!\!\uparrow\downarrow} \pm \ket{\!\!\downarrow\uparrow} \right)$. 
We set the mixing parameter to $p = 0.9$ so as to ensure a non-local input. 
${\bf D}$ is taken along $z$ axis and $B = B_0 \frac{1}{\sqrt{3}} (1,1,1)$. We 
consider FM and AFM Heisenberg exchange for each of the four input Werner 
states. Physical parameters $B_0 = 1.00 \ {\rm T}$, $|J|/B_0=1\mu_B$, 
$D/|J| = 0.4$, and $\kappa = 0.5$ are used.}
\label{fig:Concurr and non-loc}
\end{figure}

We now consider non-pure input states. As we shall see, this opens up for the 
possibility of having revival-type quantum non-locality effects. Suitable states 
to examine are of Werner type \cite{werner89}:  $\varrho_W^{\psi} = 
\frac{1-p}{4} \mathbb{1} + p\ket{\psi} \bra{\psi}$ with  $\ket{\psi}$ any maximally 
entangled spin state. We take $\varrho (0) = \varrho_W$ with $p=0.9$, which 
is sufficiently large for the input state to be nonlocal \cite{remark5}, and consider 
each of the four maximally entangled standard Bell states $\ket{\Psi_{\pm}} = 
\frac{1}{\sqrt{2}} \left( \ket{\!\!\uparrow\downarrow} \pm \ket{\!\!\downarrow\uparrow} \right)$ 
and $\ket{\Phi_{\pm}} = \frac{1}{\sqrt{2}} \left( \ket{\!\!\uparrow\uparrow} \pm 
\ket{\!\!\downarrow\downarrow} \right)$. The pairs $\ket{\Psi_{\pm}}$ and 
$\ket{\Phi_{\pm}}$ are of AFM- and FM-type, respectively. We compare q-LLG 
dynamics of FM $(J<0)$ and of AFM $(J>0)$ coupled dimers, with non-zero 
DMI ${\bf D} = D(0,0,1)$ and external magnetic field ${\bf B} = \frac{B_0}{\sqrt{3}}(1,1,1)$. 

Figure \ref{fig:Concurr and non-loc} shows the resulting non-locality $B(\varrho)$ for 
the Werner input states. The simulations confirm a revival-type behavior of non-local 
correlation in the q-LLG dynamics. This is very different from what one expects in 
Lindbladian type dynamics, where the non-locality typically is irreversibly lost after 
a finite duration \cite{bengtson16}. The origin of the effect seen in 
Fig.~\ref{fig:Concurr and non-loc} is the perfect balance between entropy production 
and loss associated with the q-LLG dynamics. We further see that the non-local 
correlation is generally more profound for AFM coupled ($J>0$) dimer. This is due 
to the fact that the lowest energy state in the AFM case is expected to have higher 
entanglement than for FM coupled spins.  

\textit{Conclusions and outlook.---} We propose a master equation that is purity-conserving 
and can be regarded as the quantum analog of the classical LLG. Although the connection 
to both LLG and q-LL \cite{wieser13} may be demonstrated for a single spin-$s$ particle, 
no equivalent comparison is in general possible for the case of several quantum spins. 
The proposed master equation is applied to a spin-$\frac{1}{2}$ dimer. In contrast to 
the classical formulation, the Bloch vector magnitude is typically not conserved, and 
depends strongly on the type of spin-spin coupling. The most striking example is an 
AFM-type product state with AFM exchange coupling, that evolves in time due to an 
external magnetic field, in which the Bloch vector becomes completely quenched, 
meanwhile the information is stored in the correlation matrix so as to create a highly 
robust resource for entanglement-based quantum information applications. Our results 
also demonstrate that a quantum description produces a dynamics that may be faster 
than that of the classical one, which might be of relevance to understand pump-probe 
experiments and experiments on ultrafast magnetization dynamics; in this context, 
thermal effects could be accounted for by introducing a stochastic noise operator, 
which models random thermal fluctuations of the environment. The evolution of mixed 
state input states reveals an unusual finite time behavior of quantum correlations 
that profoundly differs from what is typically seen in open system dynamics. To deal 
with the exponential scaling of quantum degrees of freedom, an interesting extension 
would be to develop a hybrid multi-scale method that couples a quantum based 
description of magnetism to a classical one. This will open up for several applications, 
such as for the time evolution of molecular magnets or single atoms supported on 
magnetic substrates that can be studied by using pump-probe techniques. In addition 
to its natural applications in the realm of spin dynamics,  we foresee the q-LLG 
framework as an alternative optimization method for probing the phase space in a 
problem of quantum spins.

\section*{Acknowledgements}
\vspace{-5mm}
The authors acknowledge financial support from the Knut and Alice (KAW) foundation 
through Grants No. 2018.0060,  2021.0246, and 2022.0079. Y.L. acknowledges financial 
support from the KTH-CSC scholarship agreement Grant No. 201907090094. This work 
was partially supported by the Wallenberg Initiative Materials Science for Sustainability 
(WISE) funded by the Knut and Alice Wallenberg Foundation. A.D. acknowledges 
financial support from the Knut and Alice Wallenberg (KAW) foundation through 
Grant No. 2022.0108 and from the Swedish Research Council (VR), Grant No. 
2016-05980 and Grant No. 2019-05304. D.T. acknowledges financial support from 
the Swedish Research Council (VR), Grant No. 2023-04239. O.E. acknowledges 
support from the Swedish Research Council (VR), Grant No. 641-2013-8316 
and Grant No. 2023-04899, STandUPP and European Research Council via 
Synergy Grant No. 854843.  A.B. and O.E. acknowledge support from eSSENCE. 
The computations/data handling were enabled by resources provided by the 
National Academic Infrastructure for Supercomputing in Sweden (NAISS), partially 
funded by the Swedish Research Council through grant agreement no. 2022-06725.

\newpage 

\begin{widetext}

\begin{center}
{\large{{\bf Supplemental Material for \\ \textit{Quantum analog of Landau-Lifshitz-Gilbert dynamics}}}}
\end{center}

\section{Hermiticity}

Let $\varrho (t)$ be subject to the q-LLG equation 
\begin{eqnarray}
\dot{\varrho} (t) =\frac{i}{\hbar}[\varrho (t),H] + i\kappa [\varrho (t),\dot{\varrho} (t)]. 
\label{eq:qllg}
\end{eqnarray}
We demonstrate that $\varrho (t)$ remains Hermitian for $t>0$, given it is Hermitian at $t=0$. To this end, we prove that $\varrho^{\dagger} (t) = \varrho (t) \Rightarrow \varrho^{\dagger} (t+\delta t) = \varrho (t+\delta t)$, i.e., that $\varrho^{\dagger} (t) + \dot{\varrho}^{\dagger} (t) \delta t + \mathcal{O} \left( \delta t^2 \right) = \varrho (t) + \dot{\varrho} (t) \delta t + \mathcal{O} \left( \delta t^2 \right)$, which holds if $\dot{\varrho}^{\dagger} (t) = \dot{\varrho} (t)$ since $\varrho^{\dagger} (t) = \varrho (t)$. Indeed, we may use $\varrho^{\dagger} (t) = \varrho (t)$ to find 
\begin{eqnarray}
\dot{\varrho}^{\dagger} (t) 
 & = & \underbrace{\left(\frac{i}{\hbar} 
[\varrho (t), H] \right)^{\dagger}}_{=\frac{i}{\hbar}[\varrho^{\dagger} (t),H]} + \underbrace{\left(i\kappa\left[\varrho (t) , \dot{\varrho} (t) \right] \right)^{\dagger}}_{=i\kappa\left[\varrho^{\dagger} (t),\dot{\varrho}^{\dagger} (t)\right]} = \frac{i}{\hbar}[\varrho (t),H] + i\kappa \left[ \varrho (t),\dot{\varrho}^{\dagger} (t) \right] 
\nonumber \\ 
 & = & \frac{i}{\hbar}[\varrho (t),H] + 
i\kappa \left[\varrho (t), \underbrace{\frac{i}{\hbar}[\varrho (t),H] + i\kappa \left[\varrho (t) , \frac{i}{\hbar}[\varrho (t),H] + \cdots\right]}_{=\dot{\varrho} (t)} \right] = \dot{\varrho} (t) . 
\end{eqnarray}
\qed

\section{Non-negativity of $\varrho$}
\label{sec:time-invariant-eigenvalues}

\textit{Lemma.---}Let $\varrho \equiv \varrho (t)$ be a solution of Eq.~\eqref{eq:qllg}. Then, the quantity
\begin{equation}
\label{eq:cn-constant}
C_n=\textnormal{Tr}(\varrho^n)
\end{equation}
is time-independent $\forall n\in\mathbb{N}$.

\bigskip

\textit{Proof.---}By using the property of trace, we find 
\begin{eqnarray}
\label{eq:dcn_dt}
\frac{dC_n}{dt} & = & \frac{d}{dt}\textnormal{Tr}(\varrho^n) = \textnormal{Tr}\left(\frac{d}{dt} \varrho^n \right)  = \textnormal{Tr} \left(\varrho\frac{d}{dt}\varrho^{n-1} + \varrho^{n-1} \frac{d\varrho}{dt} \right) = \ldots = \textnormal{Tr} \left(n\varrho^{n-1} \frac{d\varrho}{dt} \right) 
\nonumber \\ 
 & = & n\frac{i}{\hbar} 
\underbrace{\textnormal{Tr}\left(\varrho^{n-1}[\varrho,H]\right)}_{=0} + ni\kappa\underbrace{\textnormal{Tr}\left(\varrho^{n-1}[\varrho,\dot{\varrho}]\right)}_{=0}=0.  
\end{eqnarray}
\qed

\noindent 
Note that $n=1$ and $n=2$ are conservation of trace and purity, respectively. 

\bigskip

\textit{Theorem.---}Let $\varrho\equiv\varrho(t)\in\mathbb{C}^{N\times N}$ be a solution of Eq.~\eqref{eq:qllg} and $\lambda_j, j\in\{1,\ldots,N\}$, its real-valued eigenvalues (real-valuedness follows from Hermiticity proved above). Then, $\frac{d\lambda_j}{dt}=0$, i.e., all eigenvalues of $\varrho$ are constant in time.

\bigskip

\textit{Proof.---} Suppose that $\lambda_j$ can be divided into $G$ groups of distinct eigenvalues, each containing $M_k > 0$, $k\in\{1,\ldots,G\}$,  values, i.e., 
\begin{equation}
M_1+M_2+\ldots+M_G=N.
\end{equation}
From $\frac{dC_n}{dt}=0$ (\textit{Lemma}), we obtain
\begin{equation}
\label{eq:lambda-matrix}
\frac{dC_n}{dt}=n\sum_{k=1}^{G}\lambda_k^{n-1}M_k\frac{d\lambda_k}{dt}=0\Rightarrow\mathbf{\Lambda} \frac{d\vec{\lambda}}{dt}=0_{G\times 1},
\end{equation}
where 
\begin{eqnarray}
\mathbf{\Lambda}=\begin{pmatrix}
M_1 & M_2 & M_3 & \dots & M_G \\
M_1\lambda_1 & M_2\lambda_2 & M_3\lambda_3 & \dots & M_G\lambda_G \\
\vdots & \vdots & \vdots & \ddots & \vdots \\
M_1\lambda_1^{G-1} & M_2\lambda_2^{G-1} & M_3\lambda_3^{G-1} & \dots & M_G\lambda_G^{G-1}
\end{pmatrix}   
\end{eqnarray}
is a square $G \times G$ matrix of Vandermonde-type encompassing all powers $\{0,\ldots,G-1\}$ of the distinct eigenvalues, and 
\begin{eqnarray}
\frac{d\vec{\lambda}}{dt}=\begin{pmatrix}
\frac{d\lambda_1}{dt} \\
\frac{d\lambda_2}{dt} \\
\vdots \\
\frac{d\lambda_G}{dt}
\end{pmatrix} .
\end{eqnarray}
Now, by using the well-known determinant of a square Vandermonde matrix: 
\begin{equation}
\det{\mathbf{\Lambda}}=\left( M_1 M_2\cdots M_G \right) \cdot \left(\prod_{1\leq m<p\leq G}(\lambda_p-\lambda_m)\right)=\left(\prod_{k=1}^{G}M_k\right)\cdot\left(\prod_{1\leq m<p\leq G}(\lambda_p-\lambda_m)\right),
\end{equation}
where both factors are obviously non-zero. Thus, as $\det{\mathbf{\Lambda}}\neq0$, it follows that $\frac{d\vec{\lambda}}{dt}=0_{G\times 1}$, and the \textit{Theorem} is straightforwardly proven. \qed

According to the \textit{Theorem}, if all eigenvalues of $\varrho(0) \equiv \varrho_0$ are real and non-negative, then these eigenvalues will remain  real and non-negative for all $t > 0$. This ensures that $\varrho$ can be taken to be a valid quantum state for all $t >0$, given it is a quantum state at $t=0$. 

\section{Relation between q-LLG and q-LL dynamics}
Here, we provide details regarding the relation between the q-LLG and q-LL \cite{Wieser2013} equations. Specifically, we identify the class of states for which the two equations give rise to the same dynamics up to a rescaling of time and demonstrate that they are inequivalent for other states.

In the main text, we notice that q-LLG and q-LL are equivalent if
\begin{equation}
\label{eq:equivalence-condition}
\dot{\varrho}+\kappa^2[\varrho,[\varrho,\dot{\varrho}]]\propto\dot{\varrho}.
\end{equation}
We will show that this condition is satisfied for pure states of any number of spins and for single spins in non-pure states provided they take the form $\left( \mathbb{1} + \boldsymbol{\eta} \cdot {\bf S} \right)/(2s+1)$, but does not hold for generic non-pure states of multi-spin systems. 

First, in the pure state case, which is characterized by $\varrho^2 = \varrho$, the double commutator in Eq.~\eqref{eq:equivalence-condition} can be rewritten as 
\begin{eqnarray}
[\varrho,[\varrho,\dot{\varrho}]]=\underbrace{\varrho^2}_{=\varrho}\dot{\varrho}-2\varrho\dot{\varrho}\varrho+\dot{\varrho}\underbrace{\varrho^2}_{=\varrho} = [\varrho, \dot{\varrho}]_+ - 2\varrho\dot{\varrho}\varrho  
\end{eqnarray}
with the anti-commutator $[\varrho, \dot{\varrho}]_+ = \varrho \dot{\varrho} + \dot{\varrho} \varrho$. 
The purity assumption allows us to write  $\varrho=\ket{\psi}\bra{\psi}$, which implies 
\begin{equation}
\label{eq:part1}
\varrho\dot{\varrho}\varrho = \ket{\psi}\bra{\psi} \bra{\psi}\dot{\varrho}\ket{\psi} =  \varrho \textnormal{Tr}(\varrho\dot{\varrho}) = \varrho \frac{1}{2} \frac{d}{dt} \left(\textnormal{Tr} \varrho^2 \right) = 0 
\end{equation}
and 
\begin{eqnarray}
\label{eq:part2}
[\varrho, \dot{\varrho}]_+ & = & \ket{\psi}\bra{\psi} \left( \ket{\dot{\psi}}\bra{\psi} + \ket{\psi}\bra{\dot{\psi}} \right) + \left( \ket{\dot{\psi}}\bra{\psi} + \ket{\psi}\bra{\dot{\psi}} \right) \ket{\psi}\bra{\psi} 
\nonumber \\
 & = & \varrho 
\underbrace{\left( \langle \psi \ket{\dot{\psi}} + \langle \dot{\psi} \ket{\psi} \right)}_{=\frac{d}{dt} \langle \psi \ket{\psi} = 0} + \underbrace{\ket{\psi}\bra{\dot{\psi}} + \ket{\dot{\psi}}\bra{\psi}}_{=\dot{\varrho}} = \dot{\varrho}.
\end{eqnarray}
By combining Eqs.~\eqref{eq:part1} and \eqref{eq:part2}, we find 
\begin{equation}
\label{eq:condition-equivalence2}
\dot{\varrho}+\kappa^2[\varrho,[\varrho,\dot{\varrho}]] = \dot{\varrho} + \kappa^2\left( [\varrho,\dot{\varrho}]_{+} - 2\varrho \dot{\varrho} \varrho \right) =(1+\kappa^2)\dot{\varrho},
\end{equation}
therefore satisfying the condition of equivalence in Eq.~\eqref{eq:equivalence-condition} up to rescaling of time 
$t \mapsto (1+\kappa^2)t$, in perfect analogy with the classical 
case \cite{lakshaman84}. 

Secondly, for single spin prepared in the special form of non-pure 
state $(\mathbb{1} + \boldsymbol{\eta} \cdot {\bf S})/(2s+1)$, we find 
\begin{eqnarray}
[\varrho, [\varrho,\dot{\varrho}]] & = & \frac{1}{(2s+1)^3} [\boldsymbol{\eta} \cdot {\bf S}, [\boldsymbol{\eta} \cdot {\bf S},\dot{\boldsymbol{\eta}} \cdot {\bf S}]] = \frac{(i\hbar)^2}{(2s+1)^3} \left[ \boldsymbol{\eta} \times ( \boldsymbol{\eta} \times \dot{\boldsymbol{\eta}} )\right] \cdot {\bf S} 
\nonumber \\ 
 & = & \frac{(i\hbar)^2}{(2s+1)^3} [(\underbrace{\boldsymbol{\eta} \cdot 
\dot{\boldsymbol{\eta}}}_{=0} \boldsymbol{\eta} - |\boldsymbol{\eta}|^2 \dot{\boldsymbol{\eta}})] \cdot {\bf S} = \frac{\hbar^2 |\boldsymbol{\eta}|^2}{(2s+1)^2} \dot{\varrho} .
\end{eqnarray}
Thus,  
\begin{equation}
\label{eq:condition-equivalence3}
\dot{\varrho}+\kappa^2[\varrho,[\varrho,\dot{\varrho}]] =\left( 1+\kappa^2 \frac{\hbar^2 |\boldsymbol{\eta}|^2}{(2s+1)^2} \right) \dot{\varrho},
\end{equation}
which leads to the q-LL equation with the rescaled time $t \mapsto \left( 1+\kappa^2 \frac{\hbar^2 |\boldsymbol{\eta}|^2}{(2s+1)^2} \right) t$. 

Finally, to show that the q-LLG and q-LL equations are inequivalent in the general multi-spin case, it is sufficient to consider the case of two spin-$\frac{1}{2}$ particles 
and use the general form $\varrho = \frac{1}{4} T_{\alpha\beta} \sigma_{\alpha} \otimes 
\sigma_{\beta}$ introduced in Eq. (7) of the main text. To show that q-LLG is different from q-LL 
in this case, we prove that $[\varrho,[\varrho,\dot{T}_{xx} \sigma_{x} \otimes \sigma_{x}]]$ 
contains a non-zero term proportional to $\dot{T}_{xx} \sigma_{\alpha} \otimes \sigma_{x}$, 
say, for some $\alpha \neq x$, as it would then follow that $\kappa^2 [\varrho,[\varrho,\dot{\varrho}]]$ cannot be proportional to $\dot{\varrho}$. A straightforward calculation gives 
\begin{eqnarray}
[\varrho,[\varrho,\dot{T}_{xx} \sigma_{x} \otimes \sigma_{x}]] \propto 
\left\{ \ldots \left( T_{xy} T_{0y} + T_{xz} T_{0z} \right) \dot{T}_{xx} \right\} \sigma_0 \otimes 
\sigma_x + \ldots ,     
\end{eqnarray}
where we focus on the term with $\alpha = 0$. The factor 
$T_{xy} T_{0y} + T_{xz} T_{0z}$ is generally non-vanishing and cannot be removed 
by using the conservation of purity $T_{\alpha\beta} \dot{T}_{\alpha\beta} = 0$. 

\section{Classical limit ($s\rightarrow\infty$)} 
Here, we present a numerical analysis of some of the most divergent dynamical behaviors from the classical case, by exploring two-spin systems, each spin associated with larger values of the total spin quantum number $s$. Specifically, we examine the special cases $s=\{\frac{1}{2},1,\frac{3}{2},\ldots,5\}$. Our aim is to provide evidence for  that, even for these finite values of $s$, the behavior of the spin pair asymptotically (and slowly) approaches classical dynamics as $s\rightarrow\infty$. To facilitate our evaluation and rule out any numerical issues, we restrict our analysis to pure initial states $\varrho_0$, for which one can verify that the equation of motion expressed in Eq.~(3) of the main text has the formal solution 
\begin{equation}
\label{eq:formal-solution2}
\varrho(t)=\frac{e^{-\frac{i}{\hbar}\mathcal{H}_{\textnormal{eff}}t}\varrho_{0}e^{\frac{i}{\hbar}\mathcal{H}_{\textnormal{eff}}^{\dagger}t}}{\textnormal{Tr}\left(e^{-\frac{i}{\hbar}\mathcal{H}_{\textnormal{eff}}t}\varrho_{0}e^{\frac{i}{\hbar}\mathcal{H}_{\textnormal{eff}}^{\dagger}t}\right)} 
\end{equation}
with $\mathcal{H}_{\rm eff} = \left(\frac{1-i\kappa}{1+\kappa^2}\right)H$. 

For any spin $s$, we can calculate the expectation value $\langle \mathbf{S}\rangle$ of the spin operator $\mathbf{S}$ \textit{for each particle} of the dimer. In order to visualize the trend towards the classical regime, we consider the cases where the dynamics most differ from the classical results, namely AFM-type product states with AFM ($J>0$) and FM ($J<0$) couplings. The AFM-type initial states can be constructed by taking into account the maximum spin projections, i.e., $\varrho_0 = \ket{+ s} \bra{+ s} \otimes \ket{-s} \bra{-s}$. In Figs.~\ref{fig:afm-afm-quantum} and \ref{fig:afm-fm-quantum} we show the results for both cases. We observe that the simulations are gradually approaching classical behavior. A key indicator of this trend is the behavior of $\left|\langle \mathbf{S}\rangle\right|$. While $\left|\langle \mathbf{S}\rangle\right|$ is not yet a constant of motion, as it would be in the classical case, its rate of decrease for small time $t$ slows significantly, indicating a trend towards constancy as $s\rightarrow\infty$. Furthermore, by examining the expectation values of individual components of $\mathbf{S}$, specifically the \textit{worst} cases in terms of recovery to classical behavior (depicted in Figs.~\ref{fig:afm-afm-quantum}(b) and \ref{fig:afm-fm-quantum}(b)), we observe that the amplitude of the oscillations is trending towards the classical results. Note that the deviation between the quantum and classical oscillation rates with increasing $s$, which is particularly visible in Fig.~\ref{fig:afm-fm-quantum}, originates from the spin dependence of the precession frequency via the effective dissipation rate $\kappa/(2s+1)$ (see, e.g., Eq. (6) of the main text) and can be compensated for by performing the classical simulations with a correspondingly lower Gilbert damping $\alpha$. 

\begin{figure}[htb!]
\centering
\includegraphics[width=1.1\textwidth]{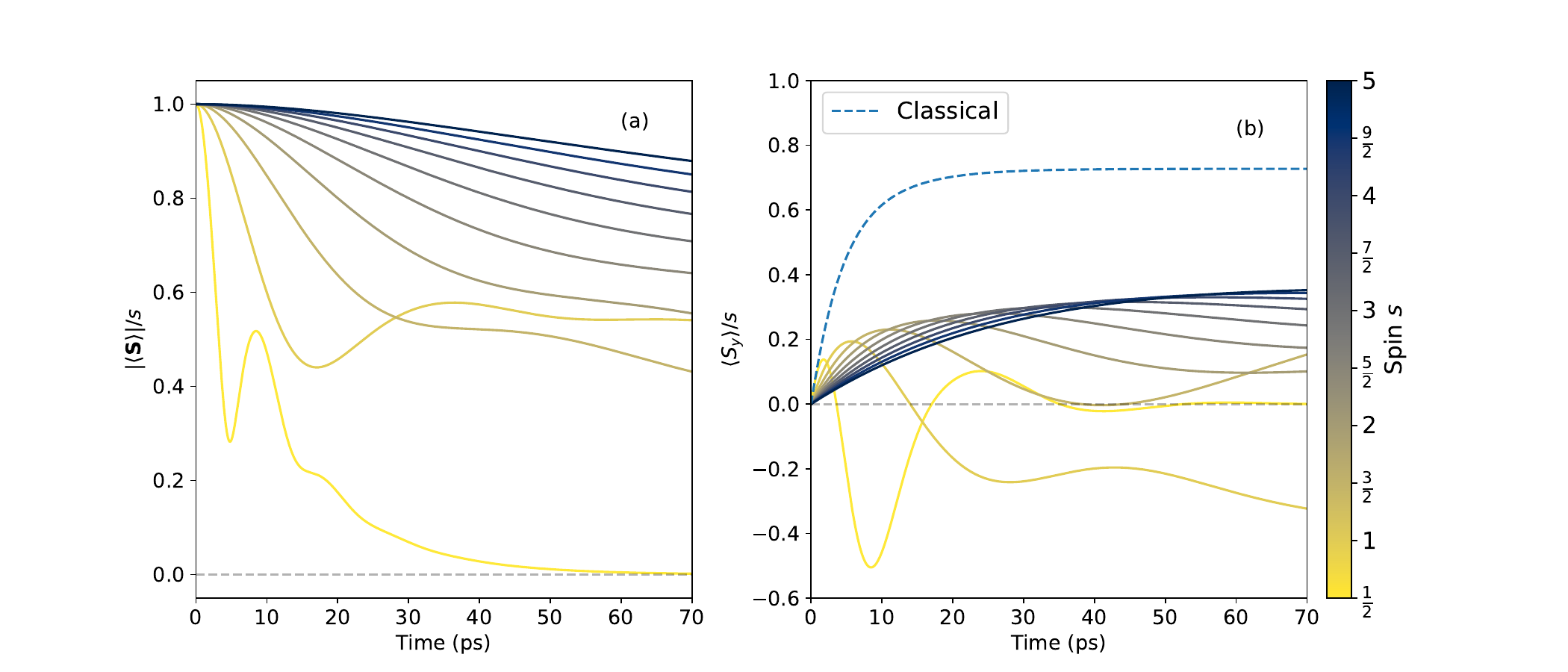}
\vspace{-4mm}
\caption{(Color online) (a) The magnitude $|\langle {\bf S} \rangle|$ of the expectation value of the spin operator $\mathbf{S}$, and (b) the expectation value $\langle S_y \rangle$ of the $y$ component of $\mathbf{S}$, for total spin quantum numbers $s=\{\frac{1}{2},1,\frac{3}{2},\ldots,5\}$. In (b) we also show the result of $m_y$ for the classical LLG simulation (with $\alpha=0.5$). In all cases, the initial state is given by an AFM-type product state $\varrho_0=\varrho_{s}\otimes\varrho_{-s}$, with an AFM coupling ($J>0$). Physical parameters $B_0 = 1.00 \ {\rm T}$, [$\mathbf{B}_0 = B_{0} (1,0,0)$], $|J|/B_0=1\mu_B$, $D/|J| = 0$, and $\kappa = 0.5$ are used.}
\label{fig:afm-afm-quantum}
\end{figure}

\begin{figure}[htb!]
\centering
\includegraphics[width=1.1\textwidth]{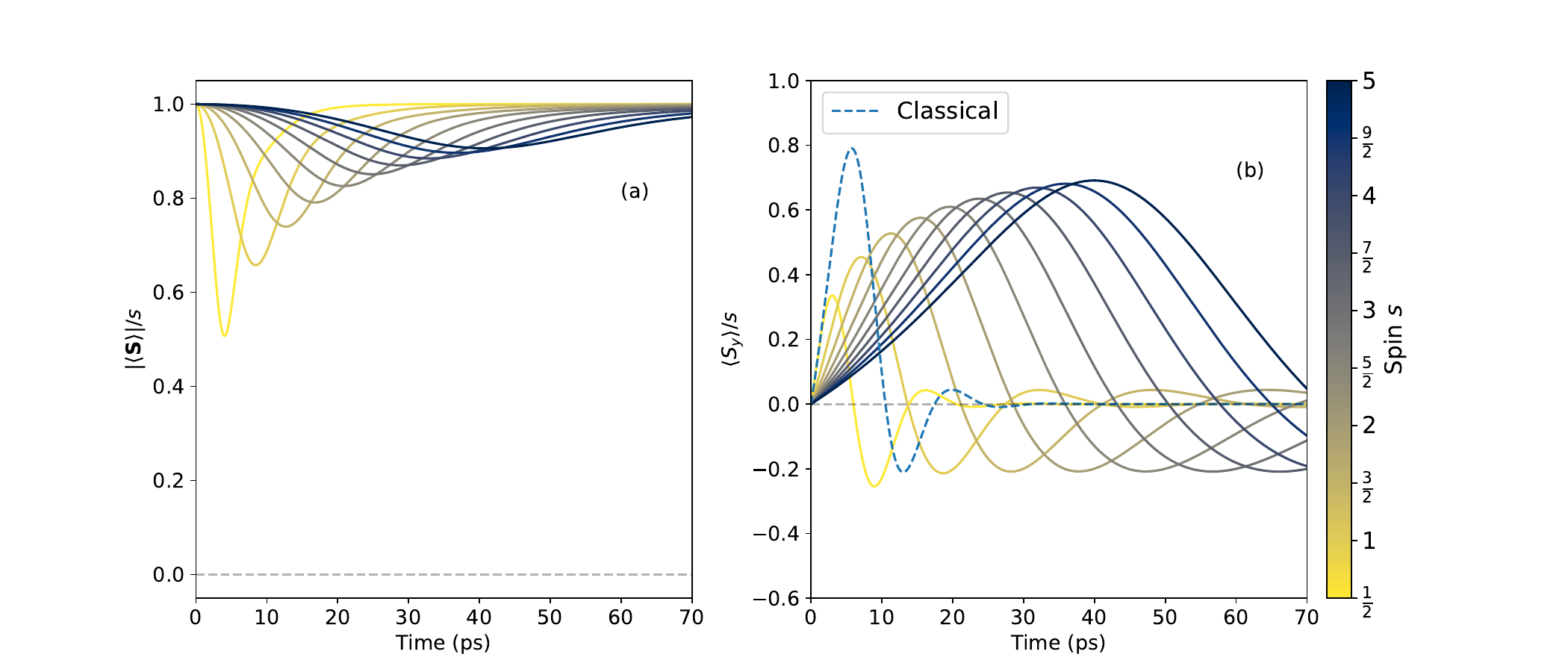}
\vspace{-4mm}
\caption{(Color online) Same as in Fig.~\ref{fig:afm-afm-quantum}, but for FM coupling ($J<0$).}
\label{fig:afm-fm-quantum}
\end{figure}

\section{Numerical methods}
Both the q-LLG and classical LLG equations are solved using a finite differences scheme (Euler's method) with time steps ranging between $10^{-3}$ and $10^{-5}$ ps. To ensure accuracy, the classical solutions are cross-verified using the more precise Runge-Kutta 4 method and the Uppsala Atomistic Spin Dynamics (UppASD) package \cite{uppasd}. Similarly, the q-LLG numerical simulations are validated against the exact solution for pure states (Eq. \ref{eq:formal-solution2}). The related Python scripts used in the present investigation can be found in GitHub \cite{numerical}.

\end{widetext}


\begin{thebibliography}{99}
\bibitem{remark1} One can note that $\boldsymbol{\eta}$ is proportional to the 
expectation value of spin, as 
Eq.~\eqref{eq:spinstate} implies 
\begin{eqnarray}
\langle {\bf S} \rangle & = & 
\frac{1}{2s+1} {\rm Tr} \left[ (\boldsymbol{\eta} \cdot {\bf S}) {\bf S} \right] = 
\frac{s(s+1)}{3} \hbar^2 \boldsymbol{\eta} , 
\label{eq:genbloch}
\end{eqnarray}
where we have used 
\begin{eqnarray}
{\rm Tr} \left( S_k S_l \right) = \frac{s(s+1)(2s+1)}{3} \hbar^2 \delta_{kl}. 
\end{eqnarray}
\bibitem{remark4} The operator space of a spin-$\frac{1}{2}$ particle is spanned 
by four operators, identity $\sigma_0$ and the three Pauli operators 
$\sigma_x,\sigma_y,\sigma_z$. This implies $4^N-1$ operators for $N$ such 
spins with the normalization condition removed.
\bibitem{remark5} A Werner state is known to be non-local if $p > \frac{1}{\sqrt{2}}$.
\end{thebibliography}
\end{document}